\documentclass[epj]{svjour}
\usepackage{graphics}
\begin{document}
\title{Surface and volume effects in the photoabsorption of nuclei 
}
\author{B. Krusche\inst{1},
J.~Lehr\inst{2},
F.~Bloch\inst{1},
M.~Kotulla\inst{1},
V.~Metag\inst{3},
U.~Mosel\inst{2},
%R.O. Owens\inst{4},
S.~Schadmand\inst{3}
}                     % Do not remove
\offprints{Bernd.Krusche@unibas.ch}          % 
\institute{Department of Physics and Astronomy, University of Basel,
           Ch-4056 Basel, Switzerland \and
	   Institut f\"ur Theoretische Physik I, Universit\"at Giessen, D-35392
           Giessen, Germany \and
	   II. Physikalisches Institut, Universit\"at Giessen, D-35392
           Giessen, Germany 
%	   \and 
%	   Department of Physics and Astronomy, University of Glasgow, Glasgow
%	   G12 8QQ, UK 
           }
\date{Received: date / Revised version: date}
\authorrunning{B. Krusche et al.}
\titlerunning{Surface and volume effects in nuclear photoabsorption}

\abstract{
Recent experimental results for meson photoproduction from nuclei 
obtained with TAPS at MAMI are analyzed 
in view of the suppression of the second nucleon resonance region 
in total photoabsorption. The cross sections can be split into a 
component from the low density surface region of nuclei 
and a component which scales more like the nuclear volume. The energy 
dependence of the surface component is similar to the deuteron cross section, 
it shows a clear signal for the second resonance peak assigned to the 
excitation of the P$_{11}$(1440), D$_{13}$(1520), and S$_{11}$(1535). The 
volume component behaves differently, it is lacking the second resonance peak 
and shows an enhancement at intermediate photon energies.   
 % end of PACS codes
} %end of abstract
\maketitle
\section{Introduction}
\label{sec:1}
The in-medium properties of hadrons is a hotly debated topic, although up to
now the experimental evidence for significant modifications of meson or
baryon properties is scarce and partly contradictory.
In-medium modifications can arise from many different effects. Undisputed are
`trivial' effects like nuclear Fermi motion, Pauli blocking of final
states or additional decay channels of nucleon resonances like $N^{\star}N$
collisions. Such effects have been investigated in particular for the 
P$_{33}$(1232) $\Delta$ resonance. More exciting perspectives are in-medium 
modifications of mesons as a consequence of partial chiral restoration 
effects. An example is the predicted shift and broadening of the $\rho$-meson 
mass distribution in the nuclear medium
\cite{Herrmann_93,Rapp_00,Cabrera_02,Post_03}, which has 
been searched for in dedicated heavy ion experiments at CERN 
(see e.g. \cite{Agakichiev_95,Adamova_03}). Such an effect would also 
influence the in-medium behavior of those nucleon resonances which have a 
significant decay branching ratio into $N\rho$.
Another much discussed effect is the in-medium modification of the $\pi\pi$
interaction in the scalar-isoscalar channel observed in pion and photon 
induced double pion production reactions
\cite{Bonutti_96,Starostin_00,Messchendorp_02}.

One of the clearest experimental observations of in-medium effects
is the complete suppression of the second resonance peak
in total photoabsorption (TPA) experiments \cite{Frommhold_94,Bianchi_94}. 
TPA on the free proton shows a peak-like structure at 
incident photon energies between 600 and 800 MeV which is attributed to the 
excitation of the P$_{11}$(1440), D$_{13}$(1520), and S$_{11}$(1535) nucleon 
resonances. This structure is not visible in nuclear TPA 
over a wide range of nuclei from lithium to uranium. 
The average over the nuclear data, normalized to the mass numbers of the 
nuclei, (`universal curve') shows only the peak of the P$_{33}$(1232) resonance
and is flat at higher incident photon energies. Many different effects have 
been invoked as an explanation. A broadening of the excitation function due 
to nuclear Fermi motion certainly contributes, but cannot explain the full
effect. Kondratyuk et al. \cite{Kondratyuk_94} and Alberico et al. 
\cite{Alberico_94} have argued for an in-medium width of the relevant nucleon 
resonances, in particular the D$_{13}$(1520), on the order of 300 MeV, i.e.
a factor of two broader than for the free nucleon. This assumption 
brings model predictions close to the data, but it is not clear what effect 
could be responsible for such a large broadening of the excited  
states \cite{Post_03,Korpa_03}. Possible effects resulting from the 
collisional broadening of the resonances have been studied in detail in the 
framework of transport models of the BUU-type (see e.g. \cite{Lehr_00}) but 
up to now the complete disappearance of the resonance structure has not been 
explained.

The resonance bump on the free proton consists of a superposition of reaction 
channels with different energy dependences
\cite{Buechler_94,Braghieri_95,Krusche_95,MacCormick_96,Haerter_97,Krusche_99},
which complicates the situation \cite{Krusche_04}. Much of the rise of the
cross section towards the maximum around 750 MeV is due to the double pion
decay channels, in particular to the  n$\pi^o\pi^+$ and p$\pi^+\pi^-$ final
states. Gomez Tejedor and Oset \cite{Gomez_96} have pointed out that for 
the latter the peaking of the cross section is related to an interference 
between the leading $\Delta$-Kroll-Rudermann term and the sequential decay 
of the D$_{13}$ resonance via D$_{13}\rightarrow\Delta\pi$. Hirata et al. 
\cite{Hirata_98} have argued that the change of this interference 
effect in the nuclear medium is one of the most important reasons for the 
suppression of the bump. Recently, an investigation of 
the reaction $\gamma p\rightarrow n\pi^o\pi^+$ has shown that the 
D$_{13}$ resonance couples strongly to the $N\rho$ decay channel
\cite{Langgaertner_01,Nacher_01}. Consequently, any shift of the 
in-medium spectral strength of the $\rho$ to lower masses would have 
a large effect on the decay width of D$_{13}\rightarrow N\rho$.  

Inclusive measurements like TPA give no further clues to effects related to 
specific reaction channels. Therefore, during the last few years, exclusive 
meson production from nuclei has been studied in the second resonance region 
\cite{Krusche_04,Roebig_96,Yorita_00,Krusche_01}. In all cases no 
significant broadening or suppression of the resonance structure 
beyond trivial nuclear effects was found in the experiments. 
On the other hand, models \cite{Post_03,Korpa_03} predict such modifications.
However, in contrast to TPA
exclusive meson production reactions are dominated by the nuclear surface. 
This is mainly due to final state interaction (FSI) of the mesons and can 
be further enhanced by density dependent partial decay widths of nucleon 
resonances \cite{Lehr_01}.  
It was found for all investigated exclusive reaction channels 
that the cross sections scale with $A^{2/3}$ ($A$ = atomic mass number), 
which indicates that due to the strong FSI only the low density 
($\rho\approx\rho_o /2$) nuclear surface contributes. This leaves 
open the possibility that the effect observed in TPA is 
related to a broadening of nucleon resonances in the nuclear volume at 
normal nuclear density $\rho_o$. In the meantime, all quasifree meson
production reactions with neutral mesons, i.e. the final states
$\eta$, $\pi^o$, $\pi^o\pi^o$, and $\pi^o\pi^{\pm}$ 
\cite{Roebig_96,Krusche_04} have been measured. Furthermore, inclusive
$X\pi^o$ production was investigated \cite{Krusche_04}. This reaction
includes not only the quasifree processes but also components
which are strongly affected by FSI like double pion
production with one pion re-absorbed in the nucleus.  
These experimental results allow for the first time to some degree
a separation of surface and volume contributions in the 
nuclear response. 

\section{Results from photoproduction of pions}
\label{sec:4}

\begin{figure}[h]
\centerline{\resizebox{0.49\textwidth}{!}{%
  \includegraphics{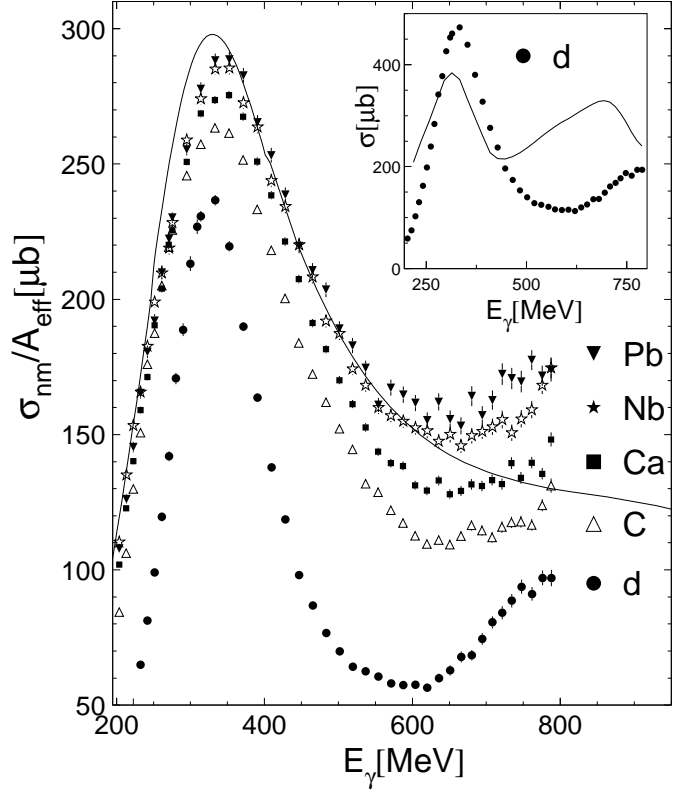}
}}
\caption{Total inclusive $\pi^o$ photoproduction cross section 
$\sigma_{nm}$ (see text). Data scaled by $A_{eff}$, $A_{eff}$=2 
for the deuteron, $A_{eff}=A^{2/3}$ for heavy nuclei. Solid curve: 
'universal curve' of TPA from nuclei scaled to  
data. Insert: $\sigma_{nm}$ (full symbols) and $\sigma_{cm}$ 
(charged mesons, solid curve) for the deuteron.
}
\label{fig:neutral}       % Give a unique label
\end{figure}

The results for the inclusive cross section for neutral meson
production $\sigma_{nm}$ on the deuteron and on heavy nuclei are summarized in 
fig. \ref{fig:neutral} \cite{Krusche_99,Krusche_04}.
Shown is the sum of the cross sections of all photoproduction
reactions with at least one $\pi^o$ or $\eta$ meson in the final state, 
with no condition on quasifree reaction kinematics. The insert compares 
for the deuteron the neutral meson production cross section $\sigma_{nm}(d)$
to the cross section $\sigma_{cm}(d)$ for pure charged meson final states 
($\pi^{\pm}$, $\pi^+\pi^-$). The latter was constructed from:
\begin{equation}
\sigma_{cm}(d) = \sigma_{abs}(d)-\sigma_{nm}(d)-\sigma_{brk}(d)
\end{equation} 
where $\sigma_{abs}(d)$ is the TPA cross section on the 
deuteron \cite{MacCormick_96} and $\sigma_{brk}(d)$ is the cross section for 
the photon induced two-body breakup of the deuteron \cite{Crawford_96}.

The comparison of the excitation functions for neutral meson production to the
`universal curve' for TPA from nuclei shows that 
the second resonance bump is much less suppressed in the meson production 
reactions. 

\begin{figure}[h]
\centerline{\resizebox{0.45\textwidth}{!}{%
  \includegraphics{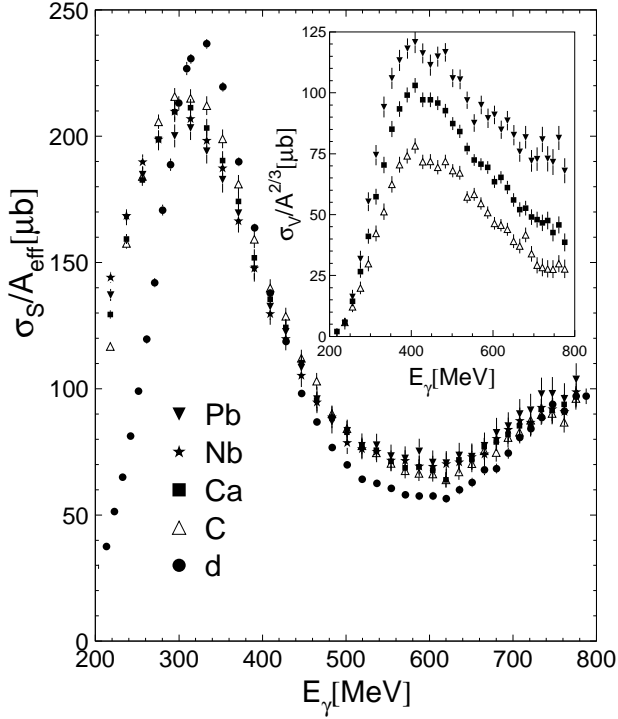}
}}
\caption{Split-up of the inclusive cross section $\sigma_{nm}$ into
the sum of quasifree exclusive partial channels $\sigma_{S}$ (main plot) 
and the non-quasifree rest (insert), $A_{eff}$ like in fig. \ref{fig:neutral} 
(see text).}
\label{fig:survol}       % Give a unique label
\end{figure}

In case of all investigated exclusive photoproduction reactions it was found in
\cite{Krusche_04} that nuclear and deuteron cross sections are to a 
good approximation related by:
\begin{equation}
\label{eq:scaling}
\frac{\sigma_{x}^{qf}(A)}{A^{2/3}}\approx\frac{\sigma_{x}^{qf}(d)}{2}
%\sigma_{x}^{qf}(A)/A^{2/3}\approx\sigma_{x}^{qf}(d)/2
\end{equation}
This is the limiting case of strong FSI effects due to the short pion mean 
free path. We define the cross section sum $\sigma_{S}$ of all quasifree 
reaction channels with neutral mesons:
\begin{equation}
\sigma_{S} = 
\sigma_{\pi^o}^{qf}+\sigma_{\eta}^{qf}
+\sigma_{2\pi^o}^{qf}+\sigma_{\pi^o\pi^{\pm}}^{qf}
\end{equation}
and the difference to the inclusive cross section:
\begin{equation}
\sigma_{V} = \sigma_{nm}-\sigma_{S}\;\;\;.
\end{equation}
Contributions from coherent single $\pi^o$ production are included into
$\sigma_{\pi^o}^{qf}$ \cite{Krusche_04} and thus also into $\sigma_{S}$.
The cross section $\sigma_{V}$ belongs to  reactions where one or more $\pi^o$
meson are produced in non-quasifree kinematics. This can be due to propagation
of the mesons in the nuclear matter, to absorption of one meson in double
pion production processes, or to production of mesons via two-body absorption
processes \cite{Carrasco_92}. The results obtained with the partial cross 
sections from 
\cite{Krusche_99,Krusche_04,Roebig_96,Krusche_95b,Weiss_03,Kleber_00} 
are summarized in fig. \ref{fig:survol}. The properly scaled quasifree cross 
sections $\sigma_{S}$ for the nuclei are very similar to the deuteron and show 
the same signal for the second resonance bump. The non-quasifree part on the 
other hand shows no indication of the second resonance peak
(for the deuteron this component vanishes of course \cite{Krusche_99}). 
The scaling of these two components with the nuclear mass number is analyzed 
with the ansatz 
\begin{equation}
\label{eq:alpha}
\sigma (A)\propto A^{\alpha}\;\;.
%$\sigma (A)\propto A^{\alpha}$.
\end{equation}  
\begin{figure}[t]
\centerline{\resizebox{0.49\textwidth}{!}{%
  \includegraphics{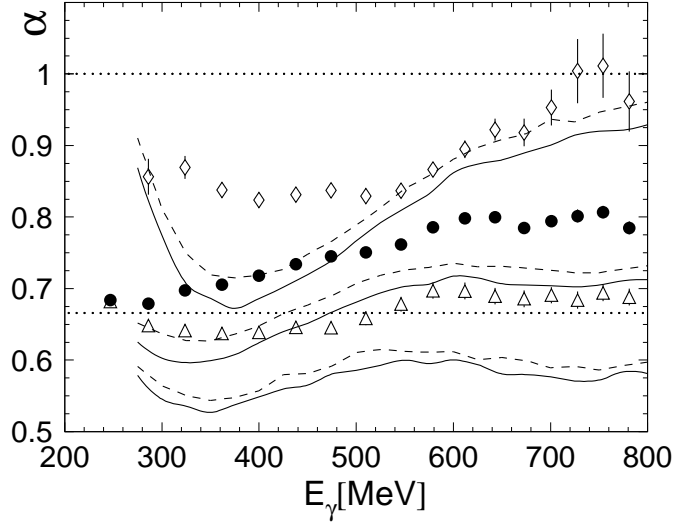}
}}
\caption{Scaling of the total cross sections with mass number. Filled circles:
inclusive cross section ($\sigma_{nm}$), open diamonds: $\sigma_V$, open
triangles: $\sigma_S$. Curves: BUU-model, solid (dashed): P$_{33}$ in-medium 
width from \cite{Hirata_79} (\cite{Oset_87}), from bottom to top corresponding
to $\sigma_S$, $\sigma_{nm}$, $\sigma_V$. 
}
\label{fig:scaling}       % Give a unique label
\end{figure}
The scaling exponents $\alpha$ are shown in
fig. \ref{fig:scaling}. The quasifree part scales like $A^{2/3}$, i.e.
like the nuclear surface. Therefore it is called $\sigma_S$.
The non-quasifree part has significantly larger
scaling coefficients, between 0.8 and unity, which indicates that this
contribution probes to some extent the nuclear volume. The corresponding
cross section is labeled $\sigma_V$. 

\begin{figure}[h]
\centerline{\resizebox{0.47\textwidth}{!}{%
  \includegraphics{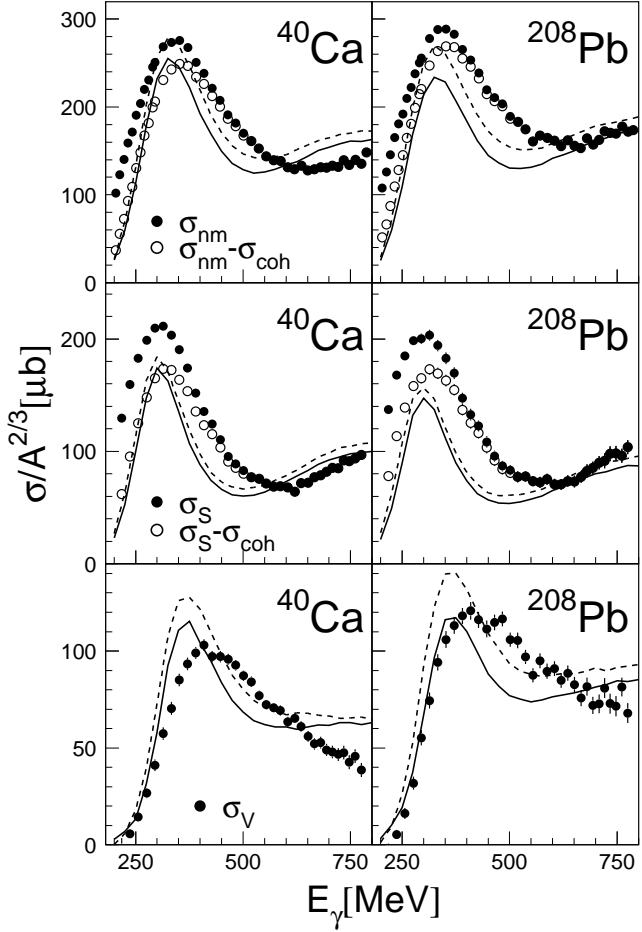}
}}
\caption{Comparison of $\sigma_{nm}$, $\sigma_{S}$ and $\sigma_{V}$ 
(full symbols) to BUU-model calculations. Open symbols: $\sigma_{nm}$,
$\sigma_{S}$ with coherent $\pi^o$ production subtracted. 
Curves: BUU-model with different prescriptions for the 
P$_{33}$(1232) in-medium width \cite{Lehr_00,Krusche_04}.
}
\label{fig:lsurvol}       % Give a unique label
\end{figure}

The data for calcium and lead are compared in fig. \ref{fig:lsurvol}
to predictions of the BUU-model (see \cite{Lehr_00,Krusche_04} for details).
The data for $\sigma_{nm}$ and $\sigma_{S}$ are also shown after subtraction of
coherent $\pi^o$ production \cite{Krusche_02} which is not included in the 
model. The magnitude of the two components is reproduced by the
model, but the resonance structures are overestimated. The model
systematically underestimates the data at photon energies between
400 and 600 MeV, which is partly due to the neglect of two-body 
absorption processes of the photon e.g. of the type $\gamma NN\rightarrow
N\Delta$ \cite{Carrasco_92}.

\section{Discussion and Conclusion}

The above results indicate a large difference between quasi\-free meson
production from the nuclear surface and non-quasifree components.
The quasifree part shows no suppression of the bump in the second
resonance region, while this is completely absent for the  
non-quasifree meson production which has larger contributions from the 
nuclear volume. 

\begin{figure}[h]
\resizebox{0.5\textwidth}{!}{%
  \includegraphics{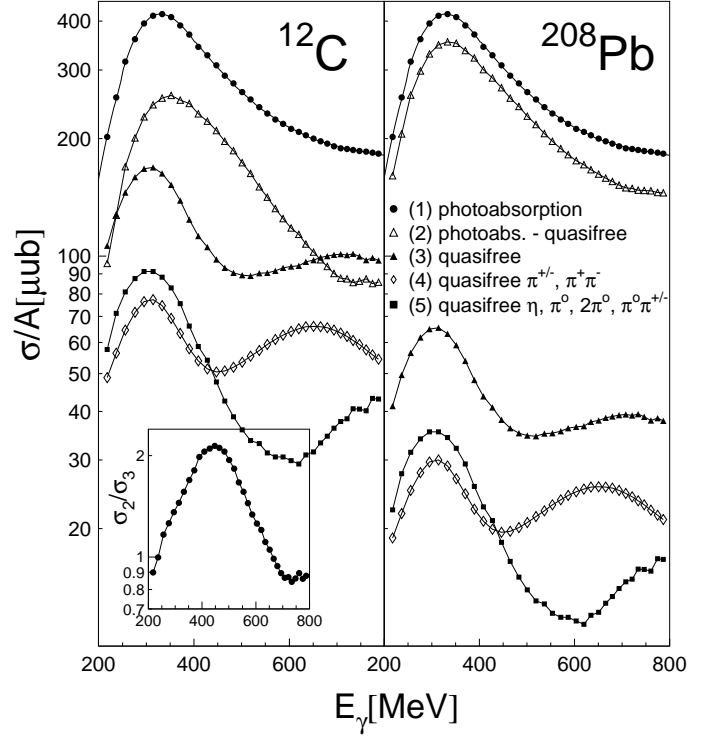}
}
\caption{Decomposition of the TPA cross section for carbon
and lead. Insert: ratio of (2) and (3) for carbon. 
}
\label{fig:splitup}       % Give a unique label
\end{figure}

There are no data available for the photoproduction of charged mesons 
from nuclei. However, since all neutral quasifree reactions follow the 
scaling eq. (\ref{eq:scaling}) and since charged pions will undergo similar 
FSI effects, it is reasonable to assume the same scaling behavior. Under 
this assumption, we can approximate the
quasifree cross section for charged meson production from the deuteron
cross section (insert in fig. \ref{fig:neutral}) with eq. (\ref{eq:scaling}).
In order to account roughly for the stronger Fermi motion effects,
the deuteron cross section was folded with a typical momentum distribution 
for nuclei. The result for carbon is shown in fig. \ref{fig:splitup} (left side, 
curve (4)) together with the quasifree cross section for neutral 
and mixed charged states ($\sigma_{nm}$ curve (5)), and the TPA
cross section  (curve (1)). The behavior for heavier nuclei 
is qualitatively the same (see fig. \ref{fig:splitup}, right side). 
The only difference is that due to the $A$-scaling of the TPA
and the $A^{2/3}$ scaling of the quasifree reactions the latter become
less important.
The sum of the quasifree meson production cross sections (curve (3)) shows
clear signals for the $\Delta$ resonance and the second resonance  
region,
although in the latter it is flatter than for the free proton.
The flattening is mainly due to Fermi motion effects.
This excitation function reflects the typical response of the
low density nuclear surface regions to photons. The difference between
this cross section and TPA represents the typical
response of the nuclear volume (curve(2)) where no isolated resonance peaks
remain. 
The insert in the figure shows the ratio of these two excitation functions
for carbon.  
The most striking feature is the buildup of strength at incident photon
energies around 400 MeV in the volume component as compared to the quasifree
surface reactions. It is known \cite{Carrasco_92} that two-body absorption 
mechanisms like $\gamma$NN$\rightarrow N\Delta$ are non-negligible in 
this energy range, but it is unknown if they alone can explain the effect. 
Further progress in the models is necessary for an understanding of this
behavior. 

\section{Acknowledgments}
We like to thank R.O. Owens for stimulating discussions.
This work was supported by Schweizerischer Nationalfonds and Deutsche 
Forschungsgemeinschaft (SFB 201).

\end{document}